\documentclass[12pt]{iopart}
\usepackage{graphicx}
\newcommand{\ba}{\begin{eqnarray}}
\newcommand{\ea}{\end{eqnarray}}
\def\be{\begin{equation}}
\def\ee{\end{equation}}

\def\vk{\mathbf{k}}

\begin{document}
\title[Pomeranchuk instabilities in lattice systems]{Pomeranchuk instabilities in multicomponent lattice systems at
finite Temperature}
\author{C.A. Lamas}
\address{Instituto de F\'\i sica de La Plata and Departamento de F\'isica, Universidad Nacional de La Plata, C.C. 67,
1900 La Plata, Argentina}
\author{D.C. Cabra}
\address{Instituto de F\'\i sica de La Plata and Departamento de F\'isica, Universidad Nacional de La Plata, C.C. 67,
1900 La Plata, Argentina}
\address{Facultad de Ingenier\'\i a, Universidad Nacional de Lomas de Zamora, Cno.\ de Cintura y Juan XXIII, (1832)
Lomas de Zamora, Argentina.}
\author{N.E. Grandi}
\address{Instituto de F\'\i sica de La Plata and Departamento de F\'isica,
Universidad Nacional de La Plata, C.C. 67, 1900 La Plata, Argentina}
\begin{abstract}
In the present paper we extend the method to detect Pomeranchuk
instabilities in lattice systems developed in previous works to study more general situations. The main result
presented here is the extension of the method to include finite temperature effects,
which allows to compute critical temperatures
 as a function of interaction strengths and density of carriers.
Furthermore, it can be applied to multiband problems which would be relevant to study systems with spin/color degrees
of freedom. Altogether, the present extended version provides a
potentially powerful technique to investigate microscopic
realistic models relevant to {\it e.g.} the  Fermi liquid to nematic transition extensively studied in connection with
different materials such as cuprates, ruthenates, etc.
\end{abstract}

\maketitle
\section{Introduction}
\label{sec:intro}

The question of the stability of the two-dimensional Landau
Fermi liquid is long-standing and of fundamental importance.
One important motivation is the non-Fermi liquid behavior observed
in high-Tc superconductors above the critical temperature, which has
in part promoted in the last years a surge of interest in possible
Pomeranchuk instabilities \cite{nematic1,nematic2,Hankevych,Metzner1,nematic3,Metzner2,Metzner3,LamasJPSJ,Hinkov,Jaku}. In
particular,
the nematic transition due to a Pomeranchuk instability in two
dimensions has been studied in detail in \cite{nematic1,Quintanilla1,Quintanilla2,nematic_MF}. For a recent review with
an
extensive list of references see \cite{LHFradkin}. Later on, similar ideas were
applied in \cite{Wuetal,Wu,YK} to include spin degrees of freedom.
In connection to ruthenates \cite{Grigeraetal,Grigera2,Arovas,Kee1}, Pomeranchuk instabilities have been advocated in
order to explain experimental observations in the metamagnetic
bilayer ruthenate Sr$_3$Ru$_2$O$_7$, where an intermediate nematic phase is
observed in a magnetic field. Another system where Pomeranchuk instabilities may play a role is in the hidden order
transition observed in URu$_2$Si$_2$ \cite{varma-2006}, in this case in connection
with spin-antisymmetric interactions.

Motivated by these observations and ideas, we have performed a
systematic study on how to analyze Pomeranchuk instabilities in lattice systems, with the aim
to account for the lattice effects, which are
expected to be strong in most of the strongly correlated electron systems \footnote{Except for the
case of the two-dimensional electron gas in strong magnetic fields}.
The effects of the lattice on the analysis of Fermi liquid instabilities have
been previously considered in {\it e.g.} \cite{Hankevych,Metzner1,Metzner3,Hinkov,nematic_MF,
Metzner_PRL,Fuseya, Frigeri,Valenzuela-2008,c2,zverev,Yama1}.

Instead of parameterizing Fermi surface (FS) deformations using
a basis of functions of the point group symmetry of the lattice, which
would be case dependent and cumbersome in general, in previous works we developed a method that is formally independent
of the underlying
lattice and straightforward to be applied \cite{nos1,nos2}.
This method can be henceforth used to detect Pomeranchuk instabilities in a generic model defined in an arbitrary
lattice,
which for completeness is reviewed in Section \ref{sec:review}.

In order to get closer to experiments, we extend the method to include finite temperature effects.
This allows to study the extended phase
diagram and in particular, to compute critical temperatures as a function of the chemical potential and interaction
strengths. On the other hand, in order to study models relevant to the systems mentioned above, we need to include
extra structure such as spin or band index, which we show how to handle
by identifying the additional degrees of freedom that parameterize spin/color flipping excitations.

The paper is organized as follows: in Section \ref{sec:review} we review the method developed  in \cite{nos1,nos2}.
Then in Section \ref{sec:extension} we present the details of the extension of the method to include finite temperature
effects. In Section \ref{sec:app} %
we show an example of application of the extended method to the finite temperature analysis of a model with an $s$-wave
instability.
In Section \ref{sec:ext_spin} we show that internal degrees of freedom, such as spin or band indexes, become manifest
as an extra matrix structure on the space of deformations, which has to be incorporated in the analysis of the
instabilities.

\section{Brief review of the spinless case at zero temperature}
\label{sec:review}
According to Landau's theory of the Fermi liquid, the change in the Landau free-energy $\Omega=E-TS-\mu N$ as a
functional of the change $\delta n(\vk)$ in the equilibrium distribution
function at finite chemical potential $\mu$ can be written, to first order
in the interaction, as
\small \ba && \!\!\!\!\!
\delta \Omega\!=\!\!\int \!\!d^{2}\!\vk\,(\varepsilon(\vk)\!-\!\mu)\,\delta n(\vk) 
\! +  \!\frac{1}{2}\!\!\int \!\!d^{2}\!\vk \!\!\int\!\!d^{2}\!\vk'
f(\vk,\vk')\,\delta n(\vk)\delta n({\vk'}) .\ \
\!\!\!\!\!\!\!\! \label{deltaE}
\ea
\normalsize
Here $\varepsilon(\vk)$ is the dispersion relation that controls the
free dynamics of the system, the interaction function $f(\vk,\vk')$
can be related to the low energy limit of the two particle vertex.
Note that we are omitting spin indices, and considering only
variations of the total number of particles
$n(\vk)=n_{\uparrow}(\vk)+n_{\downarrow}(\vk)$. This implies that
the considerations that follow in the present section will be valid
in the absence of an external magnetic field and at constant total
magnetization.

Starting from this description, Pomeranchuk has shown long time ago a simple way
to detect instabilities of the Landau Fermi liquid theory by considering the change in sign of
the energy when slightly modifying the occupation numbers \cite{Pomeranchuk}.
In his original paper he worked with a three dimensional model with a spherical FS, and hence computations could be
easily performed by using spherical harmonics.

In \cite{nos1,nos2} we have extended the original Pomeranchuk idea to deal
with lattice systems or, more generally, to systems with a Fermi surface with
an arbitrary shape where no base functions, playing the role of the spherical
harmonics in Pomeranchuk's case, can be easily identified. This problem has
been previously studied using renormalization group techniques in \cite{Metzner_PRL}
for the square lattice case with next-nearest-neighbor hopping, where it was reduced
to the diagonalization of a matrix whose rank is of the order of the number of sites.

In the present paper we generalize our approach to deal with finite temperature effects.
We also include spin degrees of freedom as well as any other internal index, such as the band
index in a multiband system.

The key idea in Pomeranchuk's analysis is to characterize
deformations $\delta n(\vk)$ that would lead to $\delta \Omega<0$, then corresponding to an instability. In order to
perform this analysis in an arbitrary lattice, the first step is to change variables in (\ref{deltaE}) from $(k_x,k_y)$
to normal Fermi variables $(g,s)$ defined as
\ba
g&=&g(\textbf{k})\equiv\mu-\varepsilon(\textbf{k})\,,\nonumber \\
s&=&s(\textbf{k})\, , \label{changeofvariables}
\ea
where the new variable $g$ is normal to the FS, which is
defined by $g=0$, and the tangent variable $s$ varies between
$-\pi$ and $\pi$. The associated Jacobian is given by
$ J^{-1}(g,s)=\left|{\partial(g,\,s)}/{\partial
\textbf{k}}\right|\,$.
We will assume that this change of variables is well defined almost everywhere in momentum space, with
the possible exception of isolated points at which the density of states may diverge (van Hove singularities).

In a stable system, the energy (\ref{deltaE}) should be positive for all $\delta n(\vk)$
that satisfy the constraint imposed by the Luttinger theorem \cite{Luttinger}
\be
\int d^2\! \vk \,\delta n(\vk) = 0\,.
\label{eq:luttinger}
\ee

Pomeranchuk's method roughly consists on exploring the space of
solutions of the Luttinger constraint (\ref{eq:luttinger}) to find
the possible $\delta n(\vk)$ that turn the energy (\ref{deltaE}) into negative values, thus pointing to an instability
of the system.

In the new variables, an arbitrary deformation can be parameterized
by $\delta g(g,s)$ and we can write $\delta n(g,s)$ at zero
temperature as
\ba
\delta n(g,s)&=&H[g +\delta g(g,s)]-H[g]\, ,
\label{eq:deltaN}
\ea
where $H$ is the unit step function. Replacing this into the
constraint (\ref{eq:luttinger}), changing the variables of
integration according to (\ref{changeofvariables}) and performing
the integral to lowest order in $\delta g$ we get
\be
\int ds J(s)\delta g(s)=0\,.
\label{contraint}
\ee
Here $J(s)=J(g,s)|_{g=0}$ and $\delta g(s) = \delta
g(g,s)|_{g=0}$. In the case in which  the FS has a nontrivial
topology, the integral would include a sum over all different
connected pieces.

We see that in order to solve the constraint, we can write an arbitrary deformation
$\delta g(s)$ as
\ba
J(s)\delta g(s)
=
\partial_s \lambda(s)\,,
\label{eq:lambda}
\ea
in terms of a free slowly varying function $\lambda(s)$.

Using the change of variables (\ref{changeofvariables}) and with the help of
eqs. (\ref{eq:deltaN}) and (\ref{eq:lambda}), we can rewrite the variation in the energy
$\delta \Omega$ to lowest order in $\delta g$ as
%
\ba
\delta \Omega&=&\!\!\int \!ds'\!\int \!ds \,\psi(s')\frac12
\left(\phantom{\frac12}\!\!\!\! J^{-1}(s)\delta(s-s') + f(s,s')
\right)\psi(s)\,,
\label{bilinear}
\ea
%
where we call  $\partial_s\lambda(s)=\psi(s)$ and
$ f(s,s')=f(g,s;g',s')|_{g=g'=0}$.

Note that the variation in the energy $\delta \Omega$ has two terms,
the first one contains the information about the form of the
unperturbed FS via $J^{-1}(s)$, while the second one encodes the
specific form of the interaction in $f(s,s')$. There is a clear
competition between the interaction function and the first term
that only depends on the geometry of the unperturbed FS.

We see in (\ref{bilinear}) that the stability condition is equivalent to the positive definiteness of a quadratic form
\ba
\delta \Omega=\langle\psi,\psi\rangle\, .
\label{eq:quaform}
\ea
for any function $\psi$ in the space of square-integrable functions defined on the Fermi surface $L_2[\mbox{FS}]$.
This was the main achievement in our previous paper: we have managed to express the
stability condition in a completely general form which is independent of the underlying geometry.

The rest of the procedure is completely technical and it is
described in detail in \cite{nos1,nos2}. The main idea is
to diagonalize this quadratic form by applying the Gram-Schmidt
orthogonalization procedure to an arbitrary starting basis of
$L_2[\mbox{FS}]$, with the pseudo scalar product defined by
(\ref{bilinear}). This results in a new orthogonal basis, that may
eventually contain functions with negative pseudonorms. In that
case, the corresponding variation on the energy (\ref{eq:quaform}) will
be negative and we say that we have found an unstable channel.

This summarizes the generalized Pomeranchuk method that can be applied to any two
dimensional model with arbitrary dispersion relation and interactions.
Some additional comments are in order: First, in some cases the solution of the constraint
that defines the variable $s$ can be extremely involved, and then an alternative route
should be followed as described in \cite{nos2}. Second, the minimal length that can
be resolved in phase space is determined by the total number of sites of the lattice.
In consequence, we do not need to consider deformations whose characteristic length is
smaller that such minimal length. Then if the starting basis is ordered in decreasing
order according to some characteristic length of the basis functions, we can limit our
procedure to a finite number of them.

It has to be stressed that in its present form, the method is not suitable to detect first
order phase transitions of the kind already predicted for some two dimensional systems \cite{Yama1}.
This is due to the fact that the energy is expanded up to second order in the variations of
the occupation numbers. Nevertheless, since our method determines sufficient conditions
for a phase transition to occur, additional first order instabilities
can only enlarge the unstable region.

\section{Finite temperature case}
\label{sec:extension}

To generalize our procedure to finite temperature, the variations in the mean occupation numbers
$\delta n(g,s)$ in (\ref{eq:deltaN})
have to be replaced by the corresponding expressions at finite temperature, namely the Heaviside
function $H$ in (\ref{eq:deltaN}) has to be replaced by the Fermi distribution $F$ at temperature $T$
\be
F(x)=\frac1{e^{-x/k_B\!T}+1}
\ee
then
\ba
\delta n(g,s)&=&F[g +\delta g(g,s)]-F[g]\, ,
\label{eq:deltaNT}
\ea
where as before $\delta g(g,s)$ is a small perturbation
parameterizing the deformation of the FS. Replacing into the constraint (\ref{eq:luttinger}), changing the variables
 of integration according to (\ref{changeofvariables}) we have
\ba
\int dg ds \; J[g,s] \left(F[g +\delta g(g,s)]-F[g]\right)&=&0
\ea
and expanding to first order in $ \delta g(g,s) $ we get
\ba
&&  \int dg\int ds \;  F_{,g}[g] J(g,s)\delta g(g,s) =0
\label{eq:contraintT}
\ea
where $F_{,g}$ denotes derivative of $F$ with the respect to $g$.
This is the finite temperature version of the constraint
(\ref{contraint}), that has to be solved in order to obtain the
independent degrees of freedom.

To obtain a solution, we note that deformations  $\delta g(g,s)$ satisfying
\ba
\int ds \;   J(g,s)\delta g(g,s)
=0
\label{eq:restricted}
\ea
constitute a particular set of solutions of (\ref{eq:contraintT}).
In what follows, we will restrict our search for unstable deformations to that particular set,
keeping in mind that more general solutions can only enlarge the unstable region of phase space.
The solution of (\ref{eq:restricted}) can be written as
\ba
 \delta g(g,s)=J^{-1}(g,s)\partial_{s}\lambda(s,g)
\label{eq:solution}
\ea

Now, replacing (\ref{eq:deltaNT}) into the energy of a perturbation (\ref{deltaE}),
expanding to second order in $\delta g(g,s)$, and using the solution (\ref{eq:solution}) we get
%
%
\ba
\!\!\!\!\!\!\!\!
 \delta \Omega\!&=&
 \!\int \!ds ds' \!\!\int\! dg dg'\,\psi(s,g) \frac{1}{2}\left(
 - g  \,J^{-1}(g,s)F_{,gg}[g]\delta(s-s')\delta(g-g')
\ + \right. \nonumber \\
&&+
\left. f(g,s;g',s')F_{,g}[g] F_{,g'}[g']
\phantom{\frac12}\!\!\!\!\right)
\psi(s',g')\ \
\label{eq:bilinearT}
\ea
%
%
where $\psi(g,s)=\partial_s\lambda(g,s)$. Note that the term linear in $\partial_s\lambda(g,s)$
vanishes by making use of (\ref{eq:restricted}). This expression is the finite temperature version of (\ref{bilinear}),
and allows us to write the energy as a temperature dependent bilinear form
\be
\delta \Omega =\langle\psi, \psi\rangle_{\!\!~_T}
\ee

The main difference with the zero temperature case is that now
both normal Fermi variables $g$ and $s$ appear in the integrals
and henceforth the rest of the procedure, {\it i.e.} the search
for negative eigenvalues to diagnose an instability, has to be
modified accordingly. More precisely, one has to write the energy
as the bilinear form (\ref{eq:bilinearT}) using the full
expressions for the Jacobian $J(g,s)$ and the interaction
$f(g,s;g',s')$. Since now the deformations $\psi(g,s)$ are
functions defined on the whole momentum plane, one has to choose
an arbitrary basis $\{\psi_n\}_{n\in N}$ now of the space of functions $L_2[{R}^2]$.

The rest of the procedure remains unchanged, and the instabilities can be diagnosed
just as before, by applying the Gram-Schmidt orthogonalization procedure to the chosen starting basis of
functions to find the modes which have a negative pseudonorm.

Even if at this point the method can be said to be complete, in practice the calculations needed may be very involved.
As already mentioned in section \ref{sec:extension}, it may be very difficult to obtain the explicit form of the
Jacobian for a given problem.
Moreover, even if the Jacobian is known, the integrals involved in eq.\ (\ref{eq:bilinearT}) may be very time-consuming
for a useful
numerical implementation. On the other hand, it may not be clear whether an instability found at finite temperature
survives in the zero
temperature limit. In order to shed light on these issues, we will go further into the details of the present
analysis.

First, note that even if the functions $F_{,g}[g]$ and $F_{,gg}[g]$ have noncompact support, they are strongly
suppressed outside
a region of width $k_BT$ around the origin. This implies that at low enough temperatures, the main contribution to the
integral (\ref{eq:bilinearT}) comes from the values
taken by the functions $J(g,s)$, $f(g,s;g',s')$ and $\psi(g,s)$ close to $g=0$. In particular, since the maximum of
the
function $F_{,g}[g]$ sits at $g=0$, it is natural to expand the
remaining factors in  powers of $g$. The interaction function can
be expanded as
\be
f(g,s;g',s')= \sum_{n,m}f_{nm}(s,s')\,g^ng'^m
\ee
while the Jacobian takes the form
\be
J^{-1}(g,s)= \sum_n J^{-1}_n(s) g^n
\label{eq:Jaroundg}
\ee
Decomposing the deformation $\psi(g,s)$ as
\be
\psi(g,s)= \sum_n \psi_n(s)g^n
\ee
we get
%
\ba
\!\!\!\!\!\!\!\!
 \delta \Omega\!&=&
 \!\sum_{n,m}\int \!ds ds' \psi_n(s)\frac{1}{2}\sum_{r,t}\left(\,(n+m+r+1)J^{-1}_r(s)F_{,g}^{n+m+r}
\delta_{rt}\delta(s-s')
\right. \nonumber \\ &&+ \left.
f_{rt}(s,s') F_{,g}^{n+r} F_{,g}^{m+t}
\phantom{\frac12}\!\!\!\!\right)
\psi_m(s')\ \
\ea
%
%
where we called $F_{,g}^n$ the $n$-th momentum of the function $F_{,g}[g]$, namely
\be
F_{,g}^n= \int dg F_{,g}[g]g^n
\ee
It has to be noted that, since $F_{,g}[g]$ is an even function, only even momenta contribute to the sum above.
Using this fact, the above expression can be rewritten as
%
\ba
\!\!\!\!\!\!\!\!
 \delta \Omega\!&=&
 \!\sum_{n,m}\int \!ds ds'
 \psi_n(s)\frac{1}{2}\sum_{j,k}\left(\,(2k+1)J^{-1}_{2k-n-m}(s)F_{,g}^{2k}\delta_{jk}\delta(s-s')
\right.  \nonumber \\&&+ \left.
f_{2k-n,2j-m}(s,s') F_{,g}^{2k} F_{,g}^{2j}
\phantom{\frac12}\!\!\!\!\right)
\psi_m(s')\ \
\ea
%
%
Since $F^{2k}_{,g}$ scales like $(k_B T)^{2k}$, at low enough temperature we can keep only the lowest orders in $k$ and
$j$ in
the formula. On the other hand, since $J^{-1}(g,s)$ and $f(g,s;g',s')$ are assumed to be regular at $g=0$, they contain
only non-negative
powers of $g$. Then in the above sum $2k-n-m \ge 0$, $2k-n \ge 0$ and $2j-m \ge 0$. We conclude that small $k$ and $j$
in turn imply small $n$ and $m$.
The meaning of this will become clear after developing explicitly the first orders in the temperature expansion.

To start with, let us focus on the simplest case, assuming that
the temperature is low enough so that we need to keep only the
$(k_BT)^0$ order, that is $k=0$. This in turn implies $n=m=0$. Then the suitable
deformations are limited to those that are independent of $g$. We
get
%
\ba
\!\!\!\!\!\!\!\!
\delta \Omega\!&=&\!
 \!\int \!ds ds' \psi_0(s)\frac{1}{2}\left(J^{-1}_{0}(s)\delta(s-s')
\!+\!
f_{00}(s,s')
\phantom{\frac12}\!\!\!\!\right)
\psi_0(s')
\ea
%
where we have used $F_{,g}^{0}=1$. We see that we have recovered
the expression (\ref{bilinear}) for the zero temperature case.

The next to leading order in temperature, $(k_BT)^2$, gives
\be
 \delta \Omega\!=
 \!\int \!ds ds' \sum_{n,m=0}^2\psi_n(s)M_{nm}(s,s')\psi_m(s')
 \label{eq:temperatureaorden2}
\ee
where $M_{nm}(s,s')$ is a $3\times3$ symmetric matrix given in
terms of the $J_n^{-1}(s)$'s and the $f_{nm}(s,s')$'s. $M$ can be
easily diagonalized to give
\ba
\delta \Omega\!&=&\!\!\int \!\!ds  ds'\left(
\tilde\psi_{0}(s)\tilde M_{00}\tilde\psi_{0}(s')
+
\tilde\psi_{1}(s)\tilde M_{11}\tilde\psi_{1}(s')
\right)\ \ \ \ \ 
\label{eq:funcional00}
\ea
where $\tilde\psi_0(s)$ and $\tilde\psi_1(s)$ are new arbitrary functions
(which are linear combinations of the original $\psi_0(s)$ and $\psi_1(s)$), and the diagonal matrix elements read
\ba
&&\!\!\!\!\!\!\!\!\!\!\!\!\!\!\!\!\!\!\tilde M_{00}=\frac12
\left( J^{-1}_{0}(s)+{\pi^{2}(kT)^{2}}J^{-1}_{2}(s) \phantom{\frac12}\!\!\!\!\right)\delta(s-s')
+\,\frac12f_{00}(s,s')
+\frac{\pi^2}{3}(kT)^2f_{02}(s,s')
\nonumber\\
&&\!\!\!\!\!\!\!\!\!\!\!\!\!\!\!\!\!\!\tilde M_{11}=\frac{\pi^{2}(kT)^2}{2}
\,J^{-1}_{0}(s)\delta(s-s')\phantom{\frac{2^2}{2^2}}
\ea
where we have consistently expanded the matrix elements to order $(kT)^2$, an we
omitted the third eigenvalue $\tilde M_{33}$ since it scales as $kT^4$. Replacing into the energy we get
\small
\ba
\nonumber &&\!\!\!\!\!\!\!\!\!\!\!\!\!\!\!\!\!\!\!\!\!\!\!\!\!\!\!\!\!\!\!\!\!\!\!\!\!\!\! \delta
\Omega\!=\!\frac12\!\!\int\! \!ds ds'
\tilde\psi_{0}(s)\!
\left(\!\left( \!J^{-1}_{0}\!(s)\!+\!{\pi^{2}(kT)^{2}}J^{-1}_{2}\!(s) \right)\delta(s\!-\!s')
+f_{00}(s,s')
+\frac{2\pi^2}{3}(kT)^2\!f_{02}(s,s')
\phantom{\frac12}\!\!\!\!\!\!
\right)\!\tilde\psi_{0}(s')
+
\nonumber\\
&&+\frac{\pi^{2}(kT)^2}2\int\!\! ds ds'
\tilde\psi_{1}(s)
\,J^{-1}_{0}\!(s)\delta(s\!-\!s') \,\tilde\psi_{1}(s')
\label{eq:funcional}
\ea
\normalsize
%
%
Note that the second line containing the $\tilde \psi_1(s)$ modes,
is proportional to the energy of a free fermion at zero temperature with a positive proportionality constant.
This observation makes evident that no instability can be
originated from those modes, and that we can safely turn them off.
Focusing on the first line that contains the $\tilde\psi_0(s)$
modes, we see that the expression is analogous to (\ref{bilinear})
but with modified Jacobian and interaction function. Then we can directly apply the
method of section \ref{sec:review} to analyze the instabilities.

Corrections to the quadratic form coming from the higher orders in the temperature expansion can be obtained
systematically in a similar way.

\section{Example: $s$-wave interaction in a square lattice at finite temperature}
\label{sec:app}

In this section we apply the formalism just developed to a model
of fermions in a square lattice with a $s$-wave interaction, whose
zero temperature case was studied before in \cite{nos1}.  The comparison with the finite temperature
case is interesting by itself, but the main aim of the present
section is to show an example of application of the general
procedure presented above.

\begin{figure}[!t]
\begin{center}
\includegraphics[width=.6\textwidth]{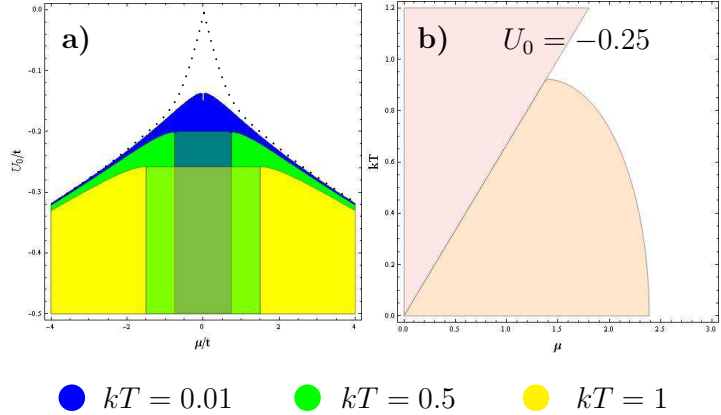}

\caption{(Color online) {\bf a)} Instability regions in the $(\mu,U_0)$ plane
for different values of $kT$. The dotted line represents the
 $T=0$ limit \cite{nos1}.  At $T=0$ the system becomes more unstable than at finite temperature, as it should be
 expected. The shaded valleys around $\mu=0$ are outside the region of validity of our low $T$
 approximation.
%
%
{\bf b)} Instability regions in the $(\mu,kT)$ plane, the boundary of the colored region defines a critical value
$T_c(\mu)$ where the instability is reached. The shaded triangular region around $\mu=0$ above the main diagonal of the
$\mu$-$kT$ plane, does not satisfy our low $T$ criterion.}
\end{center}
\label{fig:ju}
\end{figure}

We start by considering free fermions
with dispersion relation given by
\ba
\epsilon(\vk)=-2t(\cos(k_{x})+\cos(k_{y}))
\label{eq:reldisp}
\ea
At $T=0$ the FS is defined by
 \ba
  g(\vk)=\mu-\epsilon(\vk)=0
 \ea
Now we change variables according to (\ref{changeofvariables}) in the following way
\ba
g(\vk)&=&\mu+2t(\cos(k_{x})+\cos(k_{y}))\\
s(\vk)&=&\arctan \left(\frac{\tan(k_{y}/2)}{\tan(k_{x}/2)} \right)
\ea
The Jacobian of this transformation is given by
\ba
J^{-1}(g,s)=2t\sqrt{1-\beta(g\!-\!\mu)\,\cos^2(2s)}
\ea
where
\ba
\beta(x)=1-\left(\frac{x}{4t}\right)^2
\ea

This Jacobian can be expanded in powers of $g$ as in eq. (\ref{eq:Jaroundg}), obtaining
\ba
J^{-1}_{0}(s)&=&2t\sqrt{1-\beta(\mu)\cos^2(2s)}\\
J^{-1}_{1}(s)&=&-\frac{\mu \cos^{2}(2s)}{4 J^{-1}_{0}(s)}\\
J^{-1}_{2}(s)&=&\frac{\cos^{2}(2s)}{4 J^{-1}_{0}(s)}\left(1-\frac{\mu^{2} \cos^{2}(2s)}{4 {J^{-1}_{0}}^{2}(s)}\right)
\ea

In the case of the $s$-wave Pomeranchuk instability studied
in \cite{nos1}, the interaction function for the Fermi
liquid is given by the expression
\ba
f(k,k')=U_0
\label{eq:interaction}
\ea
$U_{0}$ being a constant that measures the strength of the interaction.

{
%
%
This form of the dispersion relation and interaction function can be obtained by a mean field approximation or a first
order perturbative
expansion of the Hubbard model in the square lattice, considering only hopping to nearest neighbors
\cite{Fuseya,Frigeri}
\ba
\hat{H}=-t\sum_{\sigma}\sum_{\langle
ij\rangle}\hat{c}^{\dag}_{i,\sigma}\hat{c}_{j,\sigma}+(2\pi)^2U_0\sum_{i}\hat{n}_{i\uparrow}\hat{n}_{i\downarrow}
\ea
where $\langle ij\rangle$ indicates a  sum over nearest neighbors.
}

We will expand to quadratic order in temperatures, so
that the bilinear form we have to study is given in expression
(\ref{eq:funcional00}), that specialized for the interaction
(\ref{eq:interaction}) reads
%
\be
\!\!\!\!\!\!\!\!\!\delta \Omega=\frac12\int ds ds'\;
\tilde\psi_{0}(s)
\left(\left( J^{-1}_{0}(s)+{\pi^{2}(kT)^{2}}J^{-1}_{2}(s) \right)\delta(s-s')
+{U_{0}}
\phantom{\frac12}\!\!\!\!
\right)
\tilde\psi_{0}(s')
\label{eq:funcional0}
\ee

%
%
It can be shown that higher (quartic) orders in temperature correcting the $\delta(s-s')$ coefficient in the above
expression can be safely discarded as long as $kT \ll \mu $, while the corresponding corrections proportional to the
interaction strength $U_0$ are irrelevant as long as $kT \ll t$.

As explained above, we now choose an arbitrary basis of
$L^2[\mbox{FS}]$, that in the present case will be the set of
trigonometric functions $\{\sin(ns),\cos(ns)\}$. The convenience
of this choice can be seen by the fact that each of the
coefficients in the expansion (\ref{eq:Jaroundg}) can be written
as
\ba
J^{-1}_{n}(s)=\sum_{k}j^{(k)}_{n}\cos(4ks).
\ea
Note that, if the $\cos(0s)=1$ mode has to be represented as a total derivative with respect to $s$, then
$1=\partial_s\lambda$ implies $\lambda=s$ which, being multivalued, does not belong to $L_2[FS]$. This in turn implies
that we are including a mode that does not satisfy the condition imposed by the Luttinger constraint. The deformation
of the FS originated in that mode corresponds to the addition of particles to the system.

Next we go through a Gram-Schimidt orthogonalization procedure to
obtain an orthogonal basis $\{\xi_i(s)\}$,
whose pseudo-norms $\langle\xi_i,\xi_i\rangle$ we evaluate. If for some $i$ we get a negative pseudonorm, a
deformation parameterized by $\xi_i(s)$ will have negative energy, and we can conclude that we have detected a
Pomeranchuk instability in the $i$-th channel.

We have studied the first $40$ channels detecting which ones are
unstable, and we plotted the resulting  phase diagram in the
$(\mu,kT,U_{0})$ space in Fig.1. There the
instability region determined by the dominant unstable channels is displayed.

When the temperature is increased at a constant value of the interaction $U_0$ the system becomes
more stable.
This behavior is sketched in
Fig.1a where the $(\mu,U_0)$ plane is shown for different values of $kT$,
the colored areas corresponding to unstable regions.
The phase diagram $U_0$ vs. $\mu$ of \cite{nos1} is recovered at $T=0$.

In Fig.1 we show the $U_0=-0.25$ plane. The boundary of the instability region is strongly dependent on the
chemical potential $\mu$ and we can infer the presence of an optimal value $\mu_c$ where the Fermi liquid becomes
unstable at the highest temperature.

\begin{figure}[!t]
\centering
\includegraphics[width=0.6\textwidth]{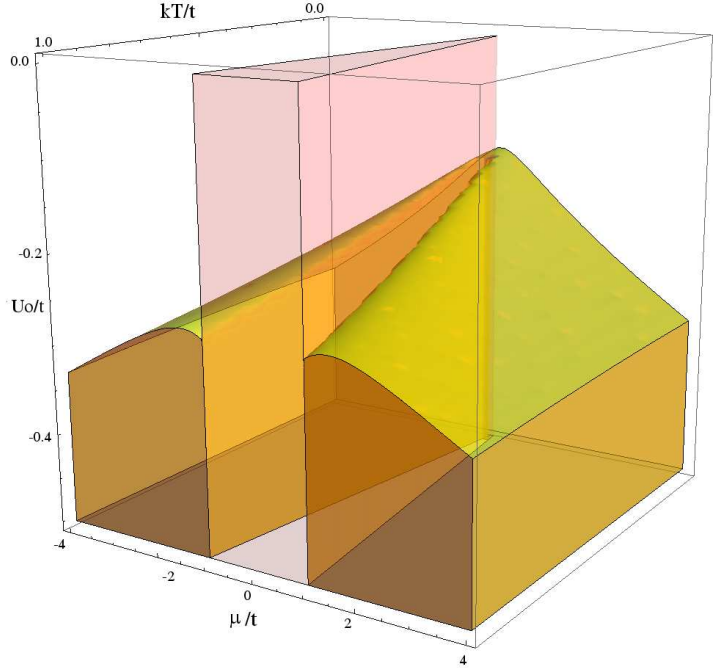}
\caption{ (Color online)  Instability region in the $(\mu,U_0,kT)$ space.
The shaded central region around the $\mu=0$ plane is outside the range of validity of our low $T$ approximation.}
\label{fig:3d}
\end{figure}

In Fig.\ref{fig:3d} we show a tree dimensional phase diagram in
 the space $(\mu,kT,U_0)$, in which the above described features can be seen in detail.

The interesting features of the phase diagram obtained by applying the method to the present model, makes evident its
potential power to study instabilities of the $d$-wave type or more complex interactions, as those discussed in
\cite{varma-2006} in the context of URu$_2$Si$_2$, etc.

We have analyzed only the first 40 modes for illustration purposes. The scaling of the unstable region as a function of
the mode index $i$ can be studied as in  \cite{nos1} and  \cite{nos2}
to show that the region of instability is bounded.



\section{Extension of the method to systems with spin/color}

\label{sec:ext_spin}

In the present section we will enlarge the approach to include the case of multiple fermionic
species (like spinful fermions, multiband materials, etc.).  Since in such materials we have one
 FS for each species, we will denote them with a new (internal) index $\alpha$, running from $1$ to $N$
 ({\it e.g.} in the spin $1/2$ case, $N=2$). Then the equations defining the FS's are given by
\be
\mu^\alpha-\varepsilon^\alpha(\textbf{k}) = 0 \hspace{0.8cm} \alpha=1,...,N
\ee

Now the constraint imposed by the Luttinger theorem on $\delta n^{\alpha}(\vk)$  reads
\be
\sum_{\alpha=1}^N \int d^2\! \vk \,\delta n^{\alpha}(\vk) = 0\,.
\label{eq:luttingers}
\ee
The changes of variables have to be done on each FS, to go from
$(k_x,k_y)$ to normal Fermi variables $(g^\alpha,s^\alpha)$:
\ba
g^\alpha&=&g^\alpha(\textbf{k})\equiv\mu^\alpha-\varepsilon^\alpha(\textbf{k})\,,\nonumber \\
s^\alpha&=&s^\alpha(\textbf{k})\, , \label{changeofvariabless}
\ea
where $g^\alpha$ is normal to the Fermi Surface corresponding to species $\alpha$
and the tangent variable $s^\alpha$ runs between $-\pi$ and $\pi$.
The associated Jacobians are given by
\be
{J_\alpha}^{\!-1}(g^\alpha,s^\alpha)=\left|\frac{\partial(g^\alpha,\,s^\alpha)}{\partial \textbf{k}}\right|\,,
\label{jacos}
\ee
and then the constraint can be rewritten as
\be
\sum_{\alpha=1}^N \int dg^\alpha ds^\alpha \,  J_\alpha (g^\alpha,\,s^\alpha)  \delta n^{\alpha}(g^\alpha,\,s^\alpha) =
0\,.
\label{eq:luttingerJ}
\ee
Now writing as before
\ba
\delta n^\alpha(g,s)&=&H[g^\alpha +\delta g^\alpha(g,s)]-H[g^\alpha]\, ,
\label{eq:deltaNs}
\ea
and replacing it into the constraint (\ref{eq:luttingerJ}) we get
\be
\sum_{\alpha=1}^N \int \!ds^\alpha \!\int_0^{\delta g^\alpha(g^\alpha\!,s^\alpha)} \!\!\!\!\!\!dg^\alpha \,  J_\alpha
(g^\alpha,\,s^\alpha)
= 0\,.
\label{eq:luttingerJs}
\ee
Performing the integrals over the $g^\alpha$'s to lowest order in $\delta g^\alpha$
\be
\sum_{\alpha=1}^N  \int ds^\alpha \,  J_\alpha (s^\alpha)
\delta g^\alpha(s^\alpha)= 0\,.
\label{constraintN}
\ee
where $J_\alpha(s^\alpha)=J_\alpha(g^\alpha,s^\alpha)|_{g^\alpha=0}$ and
 $\delta g^\alpha(s^\alpha) = \delta g^\alpha(g^\alpha,s^\alpha)|_{g^\alpha=0}$.
Being the $s^\alpha$'s dumb variables, we can use the same name on each FS and rewrite the previous expression as a
single integral over one variable $s$
\be
\int ds \, \left( \sum_{\alpha=1}^N  J_\alpha (s)
\delta g^\alpha(s)\right) = 0  \,.
\label{constraintNN}
\ee
which can be solved as before by writing
\be
\sum_{\alpha=1}^N  J_\alpha (s)
\delta g^\alpha(s) = \partial_s \lambda (s)
\label{constraintNNN}
\ee
with $\lambda(s)$ an arbitrary function of $s$.
Up to this point, we have merely added spin/color indices to the procedure outlined in
Section \ref{sec:review}, but now we note that the above expression
can be further simplified by writing

\be \delta g^\alpha(s) =
\frac1{\vert {\mathbf J} \vert^2} \left(
\partial_s \lambda(s) {J_\alpha(s)}+ J^\perp_\alpha(s)
\right)
\label{spin12}
\ee
where $\vert {\mathbf J} \vert^2 =
\sum_\alpha J_\alpha(s) J_\alpha(s)$, and $J^\perp_\alpha(s)$ is an arbitrary vector in the internal space of $\alpha$
indices that satisfy $\sum_\alpha J_\alpha^\perp(s) J_\alpha(s)=0$. This $J_\alpha^\perp(s)$ represents the new degrees
of freedom that were absent in the spinless case, that correspond to excitations that change the spin/color of the
quasiparticles. We  will need $N-1$ new arbitrary functions to parameterize them.

In the particular case of spin $1/2$, in which we will concentrate in what follows, we have $\alpha = 1,2$
corresponding to the spin up and spin down FS's respectively. Any vector perpendicular to $J_\alpha$ can be written as
\be
J^\perp_\alpha = \eta(s) \sum_\beta {\epsilon^{\alpha \beta}J_\beta}\,,
\label{spin123}
\ee
where $\epsilon^{\alpha \beta}$ is the Levi-Civita antisymmetric
symbol, and $\eta(s)$ is the new arbitrary function of $s$ needed
to parameterize the part of $\delta g^\alpha$ perpendicular to
$J_\alpha$.

Let us now compute the variation of the ground state energy corresponding to variations on the occupation numbers
$\delta n^\alpha (\vk)$
%
\be
\!\!\!\!\!\!\!\!\!\!\!\!
\!\!\!\!\!\!\!\!\!\!\!\!
\delta \Omega\!=\!
\sum_{\alpha=1}^N
\int \!\!d^{2}\!\vk\,(\varepsilon^\alpha (\vk)\!-\!\mu^\alpha)\,\delta n^\alpha (\vk)
+
\sum_{\alpha,\beta=1}^N \frac{1}{2}\!\int \!\!d^{2}\!\vk \!\!\int\! \!d^{2}\!\vk'
f^{\alpha\beta}(\vk,\vk')\;\delta n^\alpha (\vk)\delta n^\beta ({\vk'}) \, .\ \
\label{deltaEN}
\ee
%
We now change variables as in (\ref{changeofvariabless}) and express $\delta n^\alpha (\vk)$ using (\ref{eq:deltaNs})
to get
%
%
\ba
&&\!\!\!\!\!\!\!\!\!\!\!\!\!\!\!\!\!\!\!\!\!\!\!\!\!\!\!\!\!\!\!
\delta \Omega =
-\sum_{\alpha=1}^N
\int_{-\delta g^\alpha}^0 \!\!\!\!dg^\alpha \int \!\!ds^\alpha \,  J_\alpha (g^\alpha,s^\alpha) g^\alpha
\nonumber \\
&+&
\frac{1}{2}\sum_{\alpha,\beta=1}^N \!\!
\int_{-\delta g^\alpha}^0 \!\!\!\!dg^\alpha \!\!\int_{-\delta g^\alpha}^0 \!\!\!\!dg^\beta \!\!\int \!\!ds^\alpha
\!\!\int \!\!ds^\beta
f^{\alpha\beta}(g^\alpha,s^\alpha,g^\beta, s^\beta)\;J_\alpha (g^\alpha,s^\alpha) J_\beta (g^\alpha,s^\alpha) \, .\ \
\nonumber \\
\label{deltaENN}
\ea
%
Expanding to lowest order in $\delta g^\alpha$ and renaming the dumb variables $s^\alpha$ we get
%
\ba
&&\!\!\!\!\!\!\!\!\!\!\!\!\!\!\!\!\!\!\!\!\!\!\!\!
\delta \Omega=
\int \!\!ds \!\!\int \!\!ds'\!\!
\sum_{\alpha,\beta=1}^N
\delta g^\alpha (s) \,\delta g^\beta (s')
\frac{1}{2}\left(
J_\alpha (s) \delta_{\alpha \beta} \delta(s-s')
+
f^{\alpha\beta}(s,s')\;J_\alpha (s) J_\beta (s')
\right)
\label{deltaENNN}
\ea
%
where $f^{\alpha\beta}(s,s')=f^{\alpha\beta}(g^\alpha,s,g'^\beta,s')|_{g^\alpha=g'^\beta=0}$.
Other than the additional color/spin indices, the crucial difference of the above result
from the formula obtained in section \ref{sec:review} will become evident after the explicit
solution for the $\delta g^\alpha$'s which are allowed by the Luttinger theorem (\ref{spin12}) are inserted.
We then obtain
\be
\delta \Omega=
\int ds \int ds' \sum_{a,b=1,2} \psi^a(s)  K^{ab} (s,s')   \psi^b(s')\,.
\ee
Here the new indices $a,b=1,2$ are not spin/color indices, but refer to the two arbitrary
functions that parameterize the FS deformations (as in (\ref{spin12}) and (\ref{spin123})), namely
\be
\psi^1(s)= \partial_s \lambda (s)\ , \ \ \ \  \psi^2(s)=
\eta (s)
\nonumber
\ee
and the kernel $K^{ab}(s,s')$ is consequently a two-by-two matrix, which
entries are given explicitly by
\ba
&&\!\!\!\!\!\!\!\!\!\!\!\!\!\!\!\!\!\!\!\!\!\!\!\!\!\!\!\!\!\!\!\!\!\!\!\!
K^{11} (s,s') = \frac{1}{2\vert J(s)\vert^2 \vert J(s')\vert^2} \left(
\sum_\alpha J_\alpha^3(s) \delta(s-s') +\sum_{\alpha,\beta}
J_\alpha^2(s) f^{\alpha\beta}(s,s')J_\beta^2(s')
\right)
\nonumber\\
&&\!\!\!\!\!\!\!\!\!\!\!\!\!\!\!\!\!\!\!\!\!\!\!\!\!\!\!\!\!\!\!\!\!\!\!\!
K^{22} (s,s') =\frac{1}{2\vert J(s)\vert^2 \vert J(s')\vert^2} \left(
\sum_{\alpha, \beta,\gamma} \epsilon^{\alpha \beta} \epsilon^{\alpha \gamma} J_\alpha J_\beta J_\gamma \delta(s-s') \
+
\right.\nonumber \\ && \left. \ + \sum_{\alpha, \beta,\gamma,\delta} \epsilon^{\alpha\beta} \epsilon^{\gamma\delta}
J_\alpha(s) J_\beta(s) J_\gamma(s') J_\delta(s') f_{\alpha\gamma}(s,s')
\right)
\nonumber\\
&&\!\!\!\!\!\!\!\!\!\!\!\!\!\!\!\!\!\!\!\!\!\!\!\!\!\!\!\!\!\!\!\!\!\!\!\!
K^{12} (s,s') = \frac{1}{2\vert J(s)\vert^2 \vert J(s')\vert^2} \left(
\sum_{\alpha, \beta} \epsilon^{\alpha \beta} J_\alpha^2 J_\beta \delta(s-s') +
\sum_{\alpha,\gamma,\delta} \epsilon^{\gamma\delta} J_\alpha^2(s) J_\gamma(s') J_\delta(s') f^{\alpha\gamma}(s,s')
\right)
\nonumber\\
&&\!\!\!\!\!\!\!\!\!\!\!\!\!\!\!\!\!\!\!\!\!\!\!\!\!\!\!\!\!\!\!\!\!\!\!\!
K^{21} (s,s') =\frac{1}{2\vert J(s)\vert^2 \vert J(s')\vert^2} \left(
\sum_{\alpha, \beta}  \epsilon^{\alpha \beta}  J_\alpha^2 J_\beta \delta(s-s') \ +
\right.\nonumber \\ && \left. + \sum_{\alpha, \beta,\delta}
\epsilon^{\alpha\beta}
J_\alpha(s) J_\beta(s) J_\delta(s') J_\delta(s') f^{\alpha\delta}(s,s')
\right)
\label{K21}
\ea

This is the kernel to be diagonalized by the Gram-Schmidt
procedure. The main difference then is that we get an extra matrix
structure in the two dimensional space of deformations,
parallel and perpendicular to the Jacobian $J_\alpha$, and labelled by
the indices $a,b$.
This corresponds to separating those deformations that preserve total spin/color from those which do not.

The extension just presented of the generalized Pomeranchuk method can be used to study the
instabilities of the Fermi liquid under perturbations that involve spin flips, or induced by
the presence of an external magnetic field.

\section{Summary and Conclusions}
\label{sec:conclusions}

In the present work we have generalized the extended Pomeranchuk method developed
in \cite{nos1} and \cite{nos2} to detect instabilities in
more realistic lattice systems.
We have extended the method along two main directions:

First, in order to get closer to experiments, we have modified the method to include finite temperature effects. This
allows to study the extended phase diagram and in particular, to compute the critical temperature as a function of the
chemical potential and the interaction strength.

As a second direction we modified the method in order to make it applicable to study multiband systems including spin
or any other internal degree of freedom. We show that in those cases an extra matrix structure appears, parameterizing
the deformations according to whether they preserve or not the associated quantum number.
As an example of application, the effects of a finite temperature on the phase diagram of the fermion system in the
square lattice with an  $s$-wave instability were analyzed. This calculation shows the simplicity of the method and its
potentiality to identify the unstable regions of the phase diagram.

After these generalizations, the method is now ready to be applied to models relevant for different novel materials
which are thought to realize different types of Pomeranchuk transitions, such as the bilayer ruthenate
Sr$_3$Ru$_2$O$_7$, the hidden order transition material URu$_2$Si$_2$, etc. We expect to study these problems in the
near future.

\section*{ACKNOWLEDGMENTS: }
We would like to thank  E. Fradkin for helpful discussions. This
work was partially supported by the ESF grant INSTANS, PICT
ANPCYT (Grants No 20350 and 00849), and PIP CONICET (Grants No 5037, 1691 and 0396).

\newpage
\section*{REFERENCES: }
%

\end{document}